\documentclass[conference]{IEEEtran}
\IEEEoverridecommandlockouts
\usepackage{cite}
\usepackage{amsmath,amssymb,amsfonts}
\usepackage{algorithmic}
\usepackage{graphicx}
\usepackage{textcomp}
\usepackage{xcolor}
\usepackage{subfigure}
\usepackage{multirow}
\def\BibTeX{{\rm B\kern-.05em{\sc i\kern-.025em b}\kern-.08em
    T\kern-.1667em\lower.7ex\hbox{E}\kern-.125emX}}
\begin{document}


\title{Recent Results on Proportional Fair Scheduling for mmWave-based Industrial Wireless Networks\\
}

\author{\IEEEauthorblockN{Jiteng Ma}
\IEEEauthorblockA{
\text{University of Bristol}\\
Bristol, United Kingdom \\
jiteng.ma@bristol.ac.uk}
\and
\IEEEauthorblockN{Adnan Aijaz}
\IEEEauthorblockA{
\text{Toshiba Research Europe Ltd.}\\
Bristol, United Kingdom \\
adnan.aijaz@toshiba-trel.com}
\and
\IEEEauthorblockN{Mark Beach}
\IEEEauthorblockA{
\text{University of Bristol}\\
Bristol, United Kingdom \\
m.a.beach@bristol.ac.uk}
}    

\maketitle

\begin{abstract}
Millimeter wave (mmWave) communication has recently attracted significant attention from both industrial and academic communities. The large bandwidth availability as well as low interference nature of mmWave spectrum is particularly attractive for industrial communication. However, inherent challenges such as coverage and blockage of mmWave communication cause highly fluctuated channel quality. This paper explores wireless medium access control (MAC) schedulers for mmWave-based industrial wireless applications. Our objective is to design a high-performance and enhanced fairness MAC scheduling algorithm that responds rapidly to channel variations. The key contribution of our work is a method to modify the standard proportional fair (SPF) scheduler. It introduces more flexibility and dynamic properties. Compared to the SPF, our enhanced proportional fair (EPF) scheduler not only improves the priority for users in poor channel conditions but also accelerates the reaction time in fluctuated channel conditions. By providing higher fairness for all users and enhancing system robustness, it particularly adapts to the scatter-rich industrial mmWave communication environment. Through extensive performance evaluation based on the widely accepted network simulator (ns-3), we show that the new scheduler achieves better performance in terms of delivering ultra-low latency and reliable services over mmWave-based industrial communication.
\end{abstract}


\section{Introduction}
Millimeter wave (mmWave) communication is particularly attractive for industrial wireless applications due to significant bandwidth availability and low interference nature of mmWave spectrum. Novel industrial applications such as machinery precision motion control, mobile robots with collaborative operation and real-time visible monitoring have emerged in recent years\cite{brown2018ultra, pvt_5G}. These applications require real-time data, video and control signals transmission where massive throughput, ultra-fast reaction, and high reliability would be crucial. A recent study reveals that connectivity requirements for industrial control applications may require data rates in excess of 500 Mbps while demanding single-digit millisecond-level latency \cite{aijaz2018tactile}. Existing industrial wireless technologies, which are typically based on IEEE 802.15.4 and IEEE 802.11, and operate in unlicensed 2.4 GHz or 5 GHz frequency bands are not feasible for emerging industrial applications \cite{cheffena2016industrial}. 

However, mmWave communication is vulnerable in wireless environments, especially in scatter-rich industrial scenarios like factories. Small wavelength causes great pass loss and extreme signal power fluctuation in non-line-of-sight (NLOS) channels, leading to severe blockage and coverage challenges.

The overall transmission latency of a packet (denoted by \(T_\textrm{Latency}\)) is dictated by a number of factors and it can be computed as 
\begin{equation}\nonumber
\emph{T}_{\textrm{Latency}} = \emph{T}_{\textrm{Propagate}} + \emph{T}_{\textrm{Transmit}} + \emph{T}_{\textrm{PHY}} + \emph{T}_{\textrm{Queue}}
\end{equation}
where \(T_\textrm{Propagate}\) is the time for electromagnetic waves propagation over the air, \(T_\textrm{Transmit}\) is the time to position frames to the transmission time interval (TTI) at the link layer, which is a discrete value of one or multiple TTI times depending on packet size and modulation scheme, \(T_\textrm{PHY}\) is the processing time in the physical (PHY) layer, which is typically a fixed delay, and \(T_\textrm{Queue}\) is the queuing delay for buffered packets until these are allocated channel resources for transmission. Note that \(T_\textrm{Queue}\) is a variable delay which is mainly determined by the wireless medium access control (MAC) layer scheduler. Some packets need to wait for an unexpectedly long time before being assigned channel resources.

For mmWave communication, TTI values are of minimal duration owing to the implementation of ultra-high carrier frequencies. Therefore, the dominant factor in overall latency is \(T_\textrm{Queue}\). Higher latency and buffer overflows can lead to high transmission failures which are unacceptable for industrial applications. MAC layer scheduling is the key to improving the performance of mmWave communication. An effective MAC layer scheduling algorithm can fully unleash the potential of mmWave communication. To this end, the key contribution of this work is to develop a high-performance and fairness-centric MAC scheduler for scatter-rich industrial mmWave communication, which can satisfy the stringent requirement of reliability and real-time delivery for industrial applications.

The rest of the paper is structured as follows: Section II covers the required preliminaries on mmWave-related issues and the network simulator (ns-3) system module. In Section III, we discuss the limitation of some existing standard proportional fair (SPF) scheduling algorithms. Section IV presents our enhanced proportional fair (EPF) scheduler. Performance evaluation is conducted in Section V. Finally, the paper is concluded in Section IV. 

\section{Preliminaries on MmWave Communication Issues and System Module}
\subsection{MmWave Communication Issues}
MmWave communication is a state-of-the-art technology that uses frequency spectrum between 30 GHz and 300 GHz. The mmWave communication system is expected to support up to 5 Gb/s data rate for high mobility terminals and up to 50 Gb/s for static terminals \cite{bogale2016massive}. 
The large bandwidth availability is attractive for various legacy and emerging industrial application \cite{aijaz2018tactile}.
Compared to the unlicensed 2.4 GHz or 5 GHz bands, interference among different mmWave communication channels is inherently low. By leveraging beamforming technology, interference is almost entirely suppressed through spatial filtering. Besides, the size of the typical antenna is between half the wavelength and double the wavelength. Due to the short mmWave wavelength, the antenna facets are only several millimeters, allowing the deployment of a massive number of transmission and receiving antennas in portable industrial terminals. By controlling the diversity of these signals, high antenna gain is achievable.

However, mmWave communication suffers from severe propagation loss as compared to conventional communication technologies. Based on the free space pass loss (FSPL) model, the path loss value is proportional to the square of the wavelength. With the wavelength of 5 mm, the FSPL at 60 GHz is 28 decibels (dB) more than that of 2.4 GHz \cite{singh2011interference}. Besides, diffraction efficiency around the blockages is in direct proportion to the wavelength of the signal. MmWave communication channel is sensitive to obstacles, which results in significant signal scattering. This blockage problem is even worse in the industrial environment with large quantities of mechanical equipment and goods containers \cite{cheffena2016industrial}. Furthermore, in harsh industrial environments, operating heavy machines and motors equipment generates a large amount of impulsive noise. The link-level performance in terms of signal to interference plus noise ratio (SINR) is affected when adding the effects of impulsive noise on the background thermal noise. 

It has been validated that the capability of mmWave communication to diffract around obstacles is extremely weak. The NLOS channel quality suffers from high attenuation (30 dB in self-blocking condition \cite{pi2011introduction}), which is quite a tough challenge for ultra-low-latency mmWave communication. 

\subsection{MmWave Communication Module Introduction}
Network simulator ns-3 is a discrete-event tool for network researchers to develop new protocols and analyze network performance. Recently, a research group has designed a module to simulate end-to-end mmWave communication \cite{mezzavilla2018end}, which is based on the architecture of universally used long term evolution (LTE) modules \cite{baldo2011open}. The novel module includes custom channel model, PHY layer, and MAC layer. Other higher-level functions are mainly from the LTE module.

\begin{figure}
\centerline{\includegraphics[width=8.5cm]{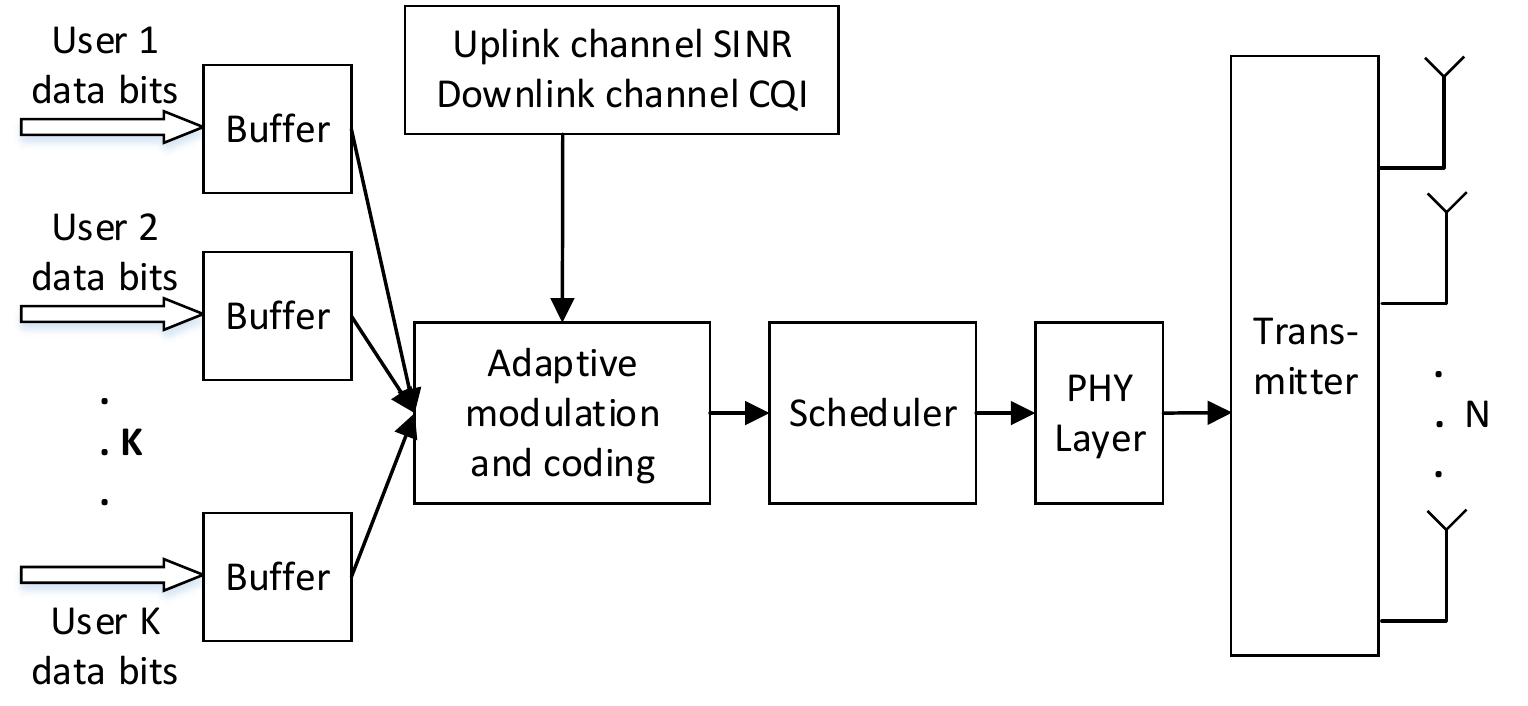}}
\caption{Transmission process at the eNodeB.}
\label{fig1}
\vspace{-0.5cm}
\end{figure}

The module implements TDD frame and subframe structure, which is similar (in principle) to the LTE system. However, compared to the fixed TTI of 1 ms in the LTE system, mmWave scheduler can allocate a flexible number of time-domain subframes to every user equipment (UE), which increases the utilization rate of the existing time slot resources \cite{ford2017achieving}. Therefore, MAC layer improves the latency and throughput performance of various users. Fig. \ref{fig1} illustrates the transmission process at evolved node base station (eNodeB).

A burst of transmitted frames first store in the buffer. This storage process triggers the scheduling mechanism. Before scheduling, SINR measurements for downlink (DL) channel and channel quality information (CQI) messages for uplink (UL) channel are collected to estimate the appropriate adaptive modulation and coding (AMC) model. The optimal modulation and coding scheme (MCS) is selected based on the AMC model and buffer state reports (BSR). The scheduler manages the transmission TTI resources. Different scheduling strategies perform differently for multi-user communication system due to the different transmission sequences for each UE. 

However, conventional scheduling algorithms focus on maximizing the system performance by ignoring some UEs with poor channel quality. For example, Max Rate scheduler gives priority to the UE with the best channel quality. When the UEs moves behind obstacles, they are suffering from extreme low priority, leading to poor communication performance. Appropriate scheduler has to adapt to mmWave fluctuated channel condition dynamically.

\section{Review of SPF Scheduling Techniques}
The SPF scheduler aims to provide an optimal trade-off between throughput and fairness. It is based on the assumption that the eNodeB receives CQI feedback from the users. The UE with the highest proportional fairness value \(Pf_k\) has priority for allocation of channel resources. The SPF scheduler performs well when the channel quality is stable. The priority for UE \(k\) in the queue is given by \eqref{eq2}
\begin{equation}
\emph{Pf}_k={\arg\max}\frac{r_k(t)}{R_k(t)},
\label{eq2}
\end{equation}
where \(r_k(t)\)  is the current data rate and  \(R_k(t)\) is the past average date rate that is derived from \eqref{eq3} as 
\begin{equation}
R_k (t)=\left\{
\begin{aligned}
&\left(1-\frac{1}{T_c}\right)R_k (t-1)+\frac{1}{T_c}r_k(t), \ \text{\(k\) is scheduled}  \\
&\left(1-\frac{1}{T_c}\right)R_k (t-1), \ \text{\(k\) is not scheduled}\label{eq3}
\end{aligned}
\right.
\end{equation}
such that \(T_c\) is the constant time window length. If UE \(k\) in the queue is scheduled to transmit frames, the average rate \(R_k(t)\) depends on the current data rate \(r_k(t)\) and past average data rate \(R_k(t-1)\). If data for UE \(k\) is waiting in the buffer, the average data rate \(R_k(t)\) drops by \((1-\frac{1}{T_c})\) compared to \(R_k(t-1)\).

However, in scatter-rich industrial environment, mmWave communication channel quality fluctuates dramatically. If a UE moves from line-of-sight (LOS) to NLOS position, the current data rate \(r_k(t)\) drops. The average rate \(R_k(t)\) drops by \((1-\frac{1}{T_c})\) , which is a very small reduction compared to the drop of \(r_k(t)\). Therefore, the priority of UE \(k\) decreases rapidly. Meanwhile, the channel quality of this NLOS UE is worse. It can only transmit via a lower modulation scheme. These negative effects cause poor latency and throughput performance of this NLOS UE. This scheduler is inefficient in the rapid fluctuated channel quality.

To remedy the limitations of the SPF scheduler, some research efforts have led to modifications of the SPF scheduling algorithm. The modified priority equation of user \(k\) is generalized as
\begin{equation}
\emph{Pf}_k={\arg\max}\frac{r_k(t)^\alpha}{R_k(t)^\beta},\label{eq4}
\end{equation}
where \(\alpha\) and \(\beta\) represent the exponential weight of the current and average data rate in the priority equation. Specifically, the equation comes to the SPF when \(\alpha=\beta=1\).

In \cite{kim2002efficient}, Kim \emph{et al.} set a fixed exponent \(\alpha\) related to the current throughput, which increases the throughput performance as well as guarantees the performance within a scope. Based on the proportional fair metric, Lee \emph{et al.} \cite{lee2009proportional} attempted to change the sequence metric to utilize the channel resources preferably, rather than simply allocating resources in turn. The approach proposed by Bechir \emph{et al.} \cite{bechir2014novel} schedules the channel resources to those previously undistributed UEs to maintain fairness. However, these modified exponents to improve fairness performance are static. They cannot adjust the strategies corresponding to the communication conditions. 

Aniba \emph{et al.} \cite{aniba2004adaptive} introduced a dynamic parameter to update the exponent \(\alpha\), which depends on discrete throughput values. The value is altered at a larger time scale to avoid rapid fluctuations. Yang \emph{et al.} proposed a method to change the exponent \(\beta\) to a  larger value that provides better fairness \cite{yang2006towards} such that \(\beta\) is derived from the past average fairness and throughput. They also proposed some schemes to improve the performance of cell-edge UEs. Xu \emph{et al.} \cite{xu2008dynamic} provide an approach to set the exponent \(\beta\) less than 1 when the UEs are in cell-edge condition. It improves the priority of cell-edge UEs. However, these approaches to update the exponent \(\alpha\) and \(\beta\) are derived from the past average parameters, which cannot match  rapid channel variations.

In \cite{yamaguchi2001forward}, Yamaguchi and Takeuchi modify the method to compute the average data rate \(R_k(t)\). A static exponent \((1-\alpha\)) is added in \eqref{eq2} to achieve high-speed data transmission. They conclude that when the \(\alpha\) is larger than 0, normally less than 1, the value of average throughput shrinks compared to SPF scheduler. The priority depends more on the current data rate. The scheduler allocates more channel resources to the terminals near the eNodeB. If the parameter \(\alpha\) is negative, the terminals far from the eNodeB get more priority. This method provides more flexibility for adjusting the priority of different UEs. However, this work only investigates the impact of static parameters.

\section{Proposed EPF Scheduler}
By reacting rapidly to the fluctuated channel quality, a practical solution is to improve the proportion of the current throughput in \eqref{eq3}, which is modified as follows.
\begin{equation}
R_k (t)=\left\{
\begin{aligned}
&\left(1-\frac{1}{T_c}\right)R_k (t-1)+\frac{1}{T_c}r_k(t)^{\gamma(t)}, \text{$k$ is scheduled}  \\
&\left(1-\frac{1}{T_c}\right)R_k (t-1), \text{$k$ is not scheduled}
\end{aligned}
\right.
\label{eq5}
\end{equation}
Note that the exponent \(\gamma(t)\) in \eqref{eq5} is related to the current channel quality. It is given by 
\begin{equation}
\gamma(t)=\frac{\delta(t)}{28}+\frac{1}{2},\label{eq6}
\end{equation}
where \(\delta(t)\) is the modulation and coding scheme (MCS) index value that dictates the  modulation scheme and coding rate as per channel conditions. The value of MCS index varies continuously from 1 to 28, in accordance with the dynamic channel quality values supported in the ns-3 mmWave module. However, \eqref{eq6} can be generalized based on MCS indices supported by a system. Introducing the effect of the channel variations leads to unevenness of throughput per UE. We set the exponent \(\gamma(t)\) as \eqref{eq6} to minimize the delay time in the scheduler. The range of exponent \(\gamma(t)\) is restricted between 0.5 and 1.5. The variation of \(\gamma(t)\) enlarges or shrinks the average throughput value for UEs in different channel condition on a reasonable scale. It enables the adjustment of average throughput value following dynamic channel variations.

\begin{figure}
\centerline{\includegraphics[width=8.5cm]{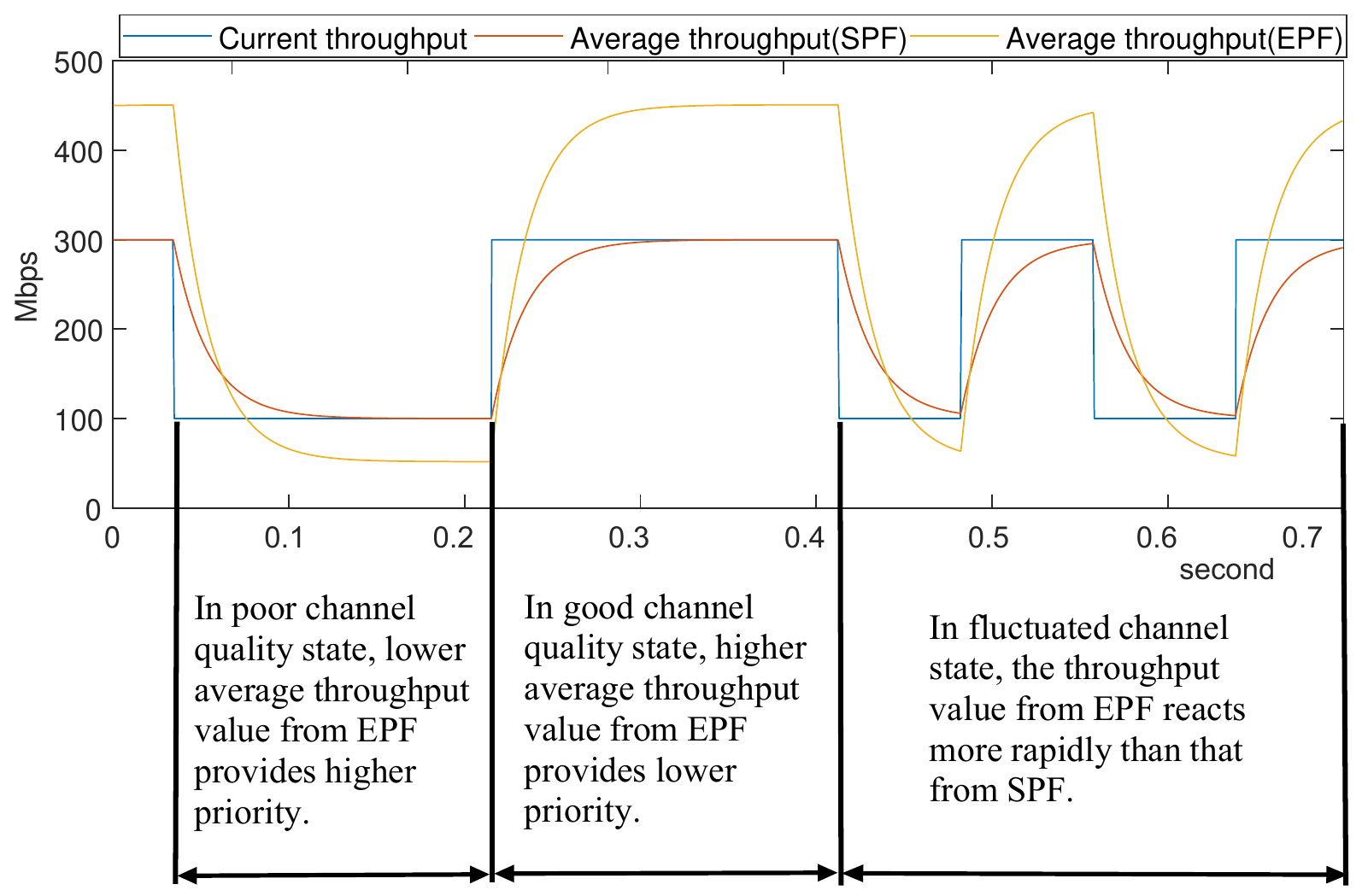}}
\caption{Average throughput variation comparison.}
\label{fig2}
\vspace{-0.5cm}
\end{figure}

Fig. \ref{fig2} compares the standard and modified average throughput values with different channel quality. In stable channel quality condition, the exponent \(\gamma(t)\) and standard average throughput value are constant. The modified average throughput value is lower than the standard value with poor channel quality state (low current throughput). When the channel quality is good (high current throughput), the modified average throughput value is larger than the standard value. Therefore, rather than SPF, EPF scheduler tends to allocate more TTI resources to the UEs who suffer from poor channel qualities. In a fast channel fluctuation condition, when a UE moves behind the obstacles, the channel quality gets worse dramatically. The instantaneous drop of \(\delta(t)\) causes a decrease of \(R_k(t)\). This NLOS UE gains high priority more rapidly in EPF. On the contrary, when the channel quality gets better, the rise of parameter \(\delta(t)\) increases the value of \(R_k(t)\), leading to a faster reduction of their priority index (\(Pf_k\)). 

Theoretically, this new EPF scheduler enjoys higher efficiency for TTI resources allocation, especially in a scatter-rich industrial communication environment. It not only improves the priority of UEs which are statically located in the NLOS position, but also accelerates the reaction time in the fluctuated channel quality scenarios.

\section{Performance Evaluation}
To fully validate the two superiorities of  new scheduler for mmWave-based communication system, we have simulated two different  cases. Case 1 has 10 moving UEs such that UE1, UE2, and UE3 are located in the scatter-rich place where channel quality fluctuates severely and the the other seven UEs are staying in LOS condition. To compare the performance limitation of different schedulers, each UE transmits 500Mbps data stream on average. This case verifies that the EPF scheduler increases the priority to the UEs with poor channel quality. Case 2 is designed to simulate realistic communication condition. The obstacles distribution, the UE location, and UE speed are all random. The average data stream for each UE is 100Mbps. Detailed simulation parameters are listed in Table \ref{tab1}.

\begin{table}[htbp]
\caption{Simulation Parameters}
\begin{tabular}{|l|l|}
\hline
\textbf{\textit{Description}}                                   & \textbf{\textit{Value}}                                   \\ \hline
Bandwidth                                                       & \(1 \times 10^9\) Hz                                      \\ \hline
Carrier frequency                                               & \(28 \times 10^9\) Hz                                     \\ \hline
Channel model type                                              & 3GPP Statistical Channel Model                            \\ \hline
Length of one subframe                                          & 100µs                                                     \\ \hline
OFDM symbols per subframe                                       & 24                                                        \\ \hline
Modulation scheme                                               & Adaptive Modulation and Coding                            \\ \hline
TTI type                                                        & Flexible TTI                                              \\ \hline
Traffic model                                                   & On-Off traffic                                            \\ \hline
RLC model                                                       & Unacknowledged                                            \\ \hline
HARQ                                                            & Enable                                                    \\ \hline
Control symbols                                                 & 1 DL and 1 UL per subframe                                \\ \hline
Payload type                                                    & UDP packet                                                \\ \hline
UE Mobility model                                               & Uniform linear motion                                     \\ \hline
\multirow{2}{*}{Traffic data stream}                              & Case 1: 500 Mbps                                          \\ \cline{2-2} 
                                                                & Case 2: 100 Mbps                                           \\ \hline
\multirow{2}{*}{UE speed}                 & Case 1: 18m/s                                                                   \\ \cline{2-2} 
                                          & Case 2: Random between 0 to 30m/s                                               \\ \hline
\multirow{2}{*}{Simulation   environment} & \begin{tabular}[c]{@{}l@{}}Case 1: 300m*300m square field, \\ 20 box obstacles,\\ eNodeB in the middle\end{tabular}           \\ \cline{2-2} 
                                          & \begin{tabular}[c]{@{}l@{}}Case 2: 300m*300m square field, \\ randomly located 300 boxes,\\ eNodeB in the middle\end{tabular} \\ \hline
\end{tabular}
\label{tab1}
\end{table}

\begin{figure}[htbp]
\centerline{\includegraphics[width=8cm]{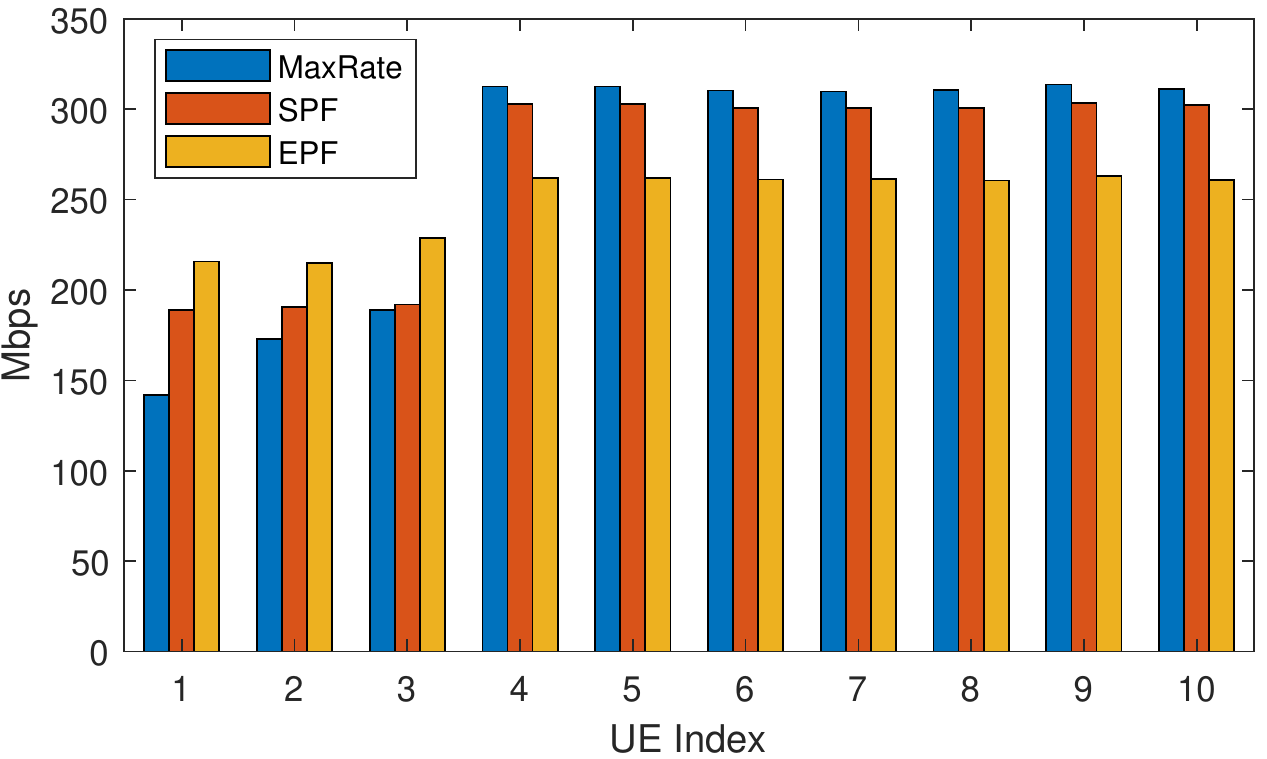}}
\caption{Case 1 throughput result comparison.}
\label{fig3}
\vspace{-0.5cm}
\end{figure}

\begin{figure}[htbp]
\centerline{\includegraphics[width=8cm]{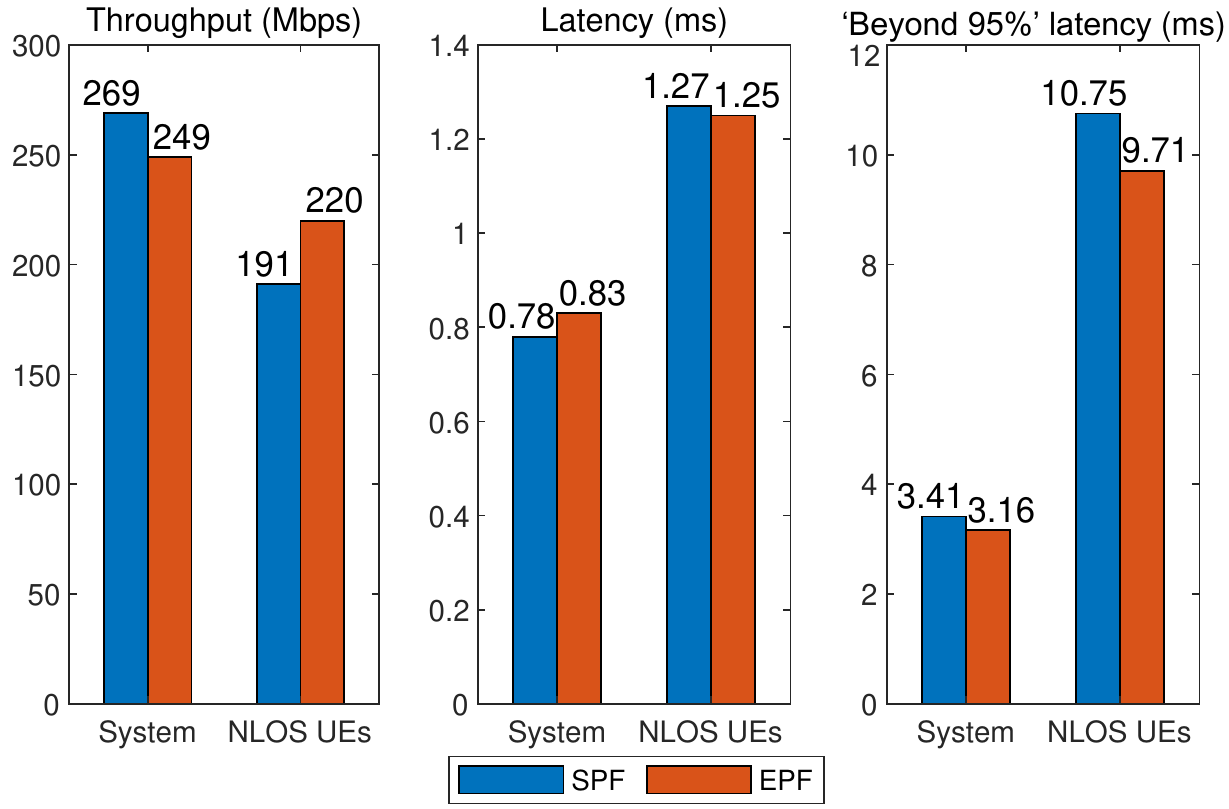}}
\caption{Case 1 system and NLOS result comparison.}
\label{fig4}
\vspace{-0.5cm}
\end{figure}

\begin{figure}[htbp]
\centerline{\includegraphics[width=8cm]{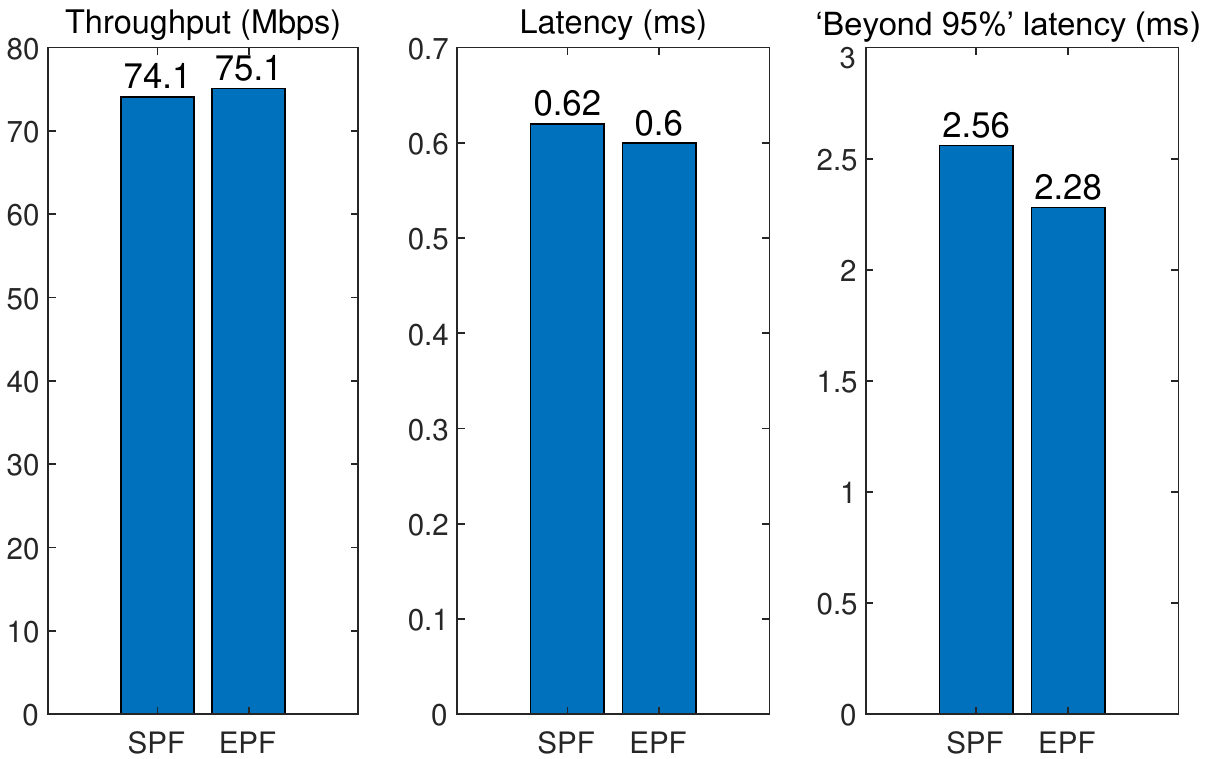}}
\caption{Case 2 system result comparison.}
\label{fig5}
\vspace{-0.5cm}
\setlength{\abovecaptionskip}{3.5cm} 
\end{figure}

\begin{figure}[htbp]
\centerline{\includegraphics[width=8cm]{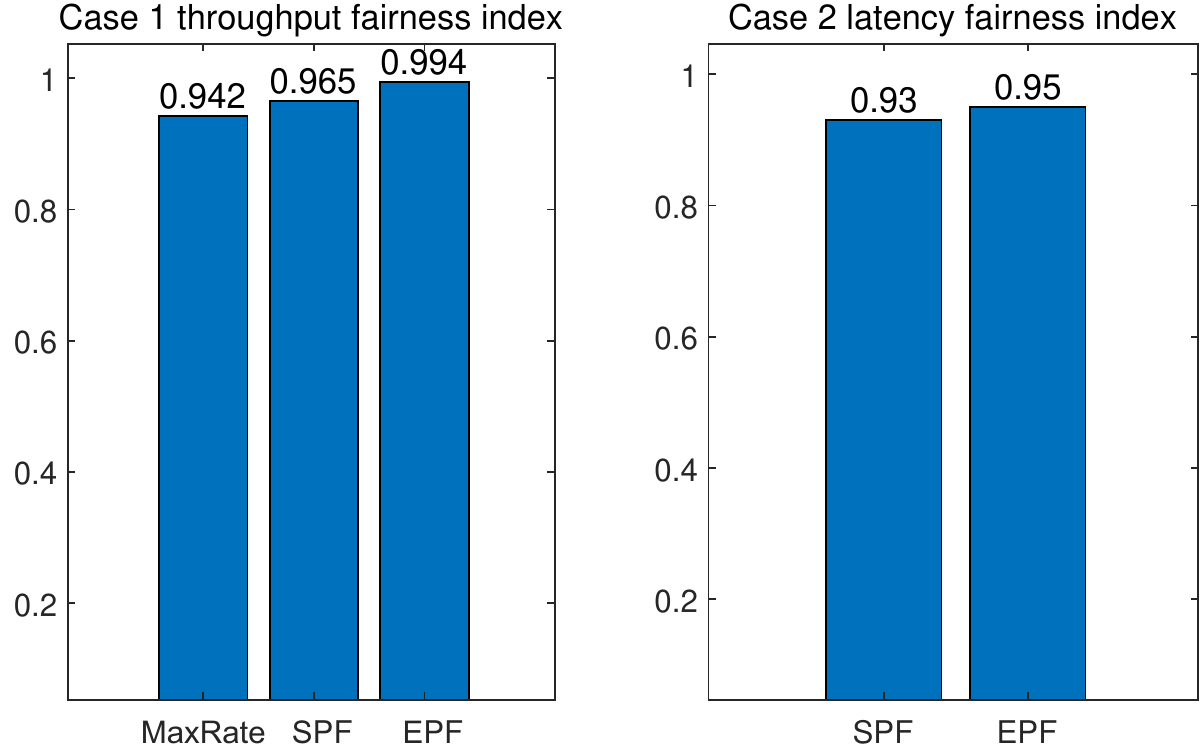}}
\caption{Case 1 throughput and case 2 latency fairness index.}
\label{fig6}
\vspace{-0.5cm}
\end{figure}

The transmitted data stream model is  chosen such that the instantaneous input data stream requirements surpass the available TTI resources, in order to trigger the scheduling process. An ``On-Off" user datagram protocol (UDP) data stream model has been used to generate  realistic data traffic. The burst stream flows for 5 µs, before the data stream stops for a period that follows the exponential distribution with an average of 100 µs. With this data stream model, the scheduler cannot satisfy the requirement of all the UEs during the burst period, resulting in backlogged queues and longer latency. SPF and EPF schedulers allocate limited TTI resources to different UEs. The simulation results are shown from Fig. \ref{fig3} -- Fig. \ref{fig6}.

We use Jain's fairness index \cite{jain1984quantitative} to compare the fairness performance of throughput and latency for two schedulers, which is calculated as 
\begin{equation}
F(K)=\frac{(\sum_{i=1}^{K} x_i)^2}{K\sum_{i=1}^{K} x_i^2},
\label{eq7}
\end{equation}
where $K$ is the number of the UEs, $x_i$ is the average throughput or latency value of the \(i\)th UE.

The parameter ‘beyond $95^{th}$ percentile’ latency is derived by sequencing the latency value of all the packets and calculate the average value of the top 5\% packets latency. When the UEs moves from LOS to NLOS channel condition, the latency result is very high.   ‘Beyond $95^{th}$ percentile’ latency value can quantify the reaction speed of channel variation.

In case 1, three NLOS UEs are moving through a scatter-rich area, where channel quality is fluctuated rapidly. The other seven UEs are staying at LOS area. From the throughput result in Fig. \ref{fig3}, max rate scheduler allocates most of the resources to the LOS UEs to maximise the overall performance by declining the NLOS priority. The proposed EPF scheduler support highest throughput performance for NLOS UEs among these three schedulers, while the trade-off is the lower performance for the rest of LOS UEs. the throughput fairness index in Fig. \ref{fig6} reaches up to 0.994 with EPF scheduler.

From the comparison between system and NLOS UEs result in Fig. \ref{fig4}, the NLOS UEs enjoys higher throughput and latency performance in EPF scheduler by compromising the system performance. Due to better throughput result of NLOS UEs, their latency performance does not improve obviously. However, the EPF ‘beyond $95^{th}$ percentile’ latency results in both system and NLOS UEs are 10.7\% better than SPF due to their fast reaction to the channel fluctuation.

Case 2 is designed to simulate a realistic random communication condition. The simulation result is shown in Fig. \ref{fig5}. The overall EPF performance is slightly higher than SPF. Particularly, the ‘beyond $95^{th}$ percentile’ latency result improve by 12.3\% due to EPF scheduler faster reaction to the channel fluctuation. The EPF latency fairness index shown in Fig. \ref{fig6} also improves 2.2\% compared with SPF.

\section{Conclusion}

This paper has proposed an EPF scheduling algorithm to overcome the limitations of SPF scheduling in dynamic wireless channels. Extensive simulation results based on mmWave communication module in ns-3 demonstrate that the proposed EPF scheduler improves the priority of UEs in the NLOS positions as compared to the SPF scheduler. EPF scheduler has better throughput and fairness performance for NLOS UEs. The trade-off results is a slight performance degradation for the LOS UEs. In realistic random communication environment, the  ‘beyond $95^{th}$ percentile’ latency drops by 12.3\%. Overall, the EPF scheduler provides an attractive solution for scatter-rich industrial environments. 

\vspace{12pt}
\bibliographystyle{IEEEtran}
\bibliography{bibliography.bib}
\end{document}